\documentclass{article}

\usepackage[final]{neurips_2024}

\usepackage[utf8]{inputenc} 
\usepackage[T1]{fontenc}    
\usepackage{hyperref}       
\usepackage{url}            
\usepackage{booktabs}       
\usepackage{amsfonts}       
\usepackage{nicefrac}       
\usepackage{microtype}      
\usepackage{xcolor}         

\usepackage{csquotes}
\usepackage{graphicx}
\title{Revisiting the Robust Alignment of Circuit Breakers}

\author{%
Leo Schwinn \\
Technical University of Munich \\ Munich Data Science Institute
\And 
Simon Geisler \\
Technical University of Munich \\ Munich Data Science Institute
}

\begin{document}

\maketitle

\begin{abstract}
  Over the past decade, adversarial training has emerged as one of the few reliable methods for enhancing model robustness against adversarial attacks~\citep{szegedy_intriguing_2014, madry_towards_2018, xhonneux2024efficient}, while many alternative approaches have failed to withstand rigorous subsequent evaluations. Recently, an alternative defense mechanism, namely \enquote{circuit breakers}~\citep{zou2024improving}, has shown promising results for aligning LLMs. 
  In this report, we show that the robustness claims of \enquote{Improving Alignment and Robustness with Circuit Breakers} against unconstraint continuous attacks in the embedding space of the input tokens may be overestimated~\citep{zou2024improving}. Specifically, we demonstrate that by implementing a few simple changes to embedding space attacks~\citep{schwinn2024adversarial, schwinn2024soft}, we achieve  $100\%$ attack success rate (ASR) against circuit breaker models. Without conducting any further hyperparameter tuning, these adjustments increase the ASR by more than $80\%$ compared to the original evaluation. 
  Code is accessible at: \url{https://github.com/SchwinnL/circuit-breakers-eval}
\end{abstract}

Neural networks are still vulnerable to adversarial examples. A recurring issue has slowed the field's progress from early work in computer vision to recent evaluations in LLMs: overly optimistic robustness assessments of new defenses. When these evaluations prove faulty, it can create a domino effect of problems. Researchers spend a considerable amount of time rectifying flawed methods~\citep{carlini_defensive_2016, carlini_adversarial_2017, athalye_robustness_2018, tramer_adaptive_2020}, while end-users may unknowingly rely on inadequate protections~\citep{hönig2024adversarialperturbationsreliablyprotect}.\citet{schwinn2024adversarial} highlight that inadequate defense evaluations have even more severe consequences in the context of LLMs due to the increased cost of performing evaluations and the potentially high real-world impact of toxic models.

Best practices for robustness evaluation have been outlined in prior work and can be useful to identify unusually high robustness claims~\citep{athalye_robustness_2018, tramer_adaptive_2020}. In the case of circuit breakers, we observe that the authors claim to achieve high robustness against unconstrained attacks in the continuous embedding space of the model~\citep{zou2024improving}. This contradicts previous findings in other domains that unconstrained continuous attacks should achieve $100\%$ attack success rate. While these findings might not trivially transfer to the alignment threat models in LLMs they still suggest that the robustness claims might be overestimated.

\section*{Experiments and Results}

We apply three simple modifications to the embedding space attack used in the original evaluation of circuit breaker models. These adjustments are sufficient to decrease the robustness to $0\%$. Notably, we implemented these changes without conducting any hyperparameter tuning.

\textbf{Optimizer.} We observe that the embedding attack used in the circuit breaker paper is very sensitive to the learning rate hyperparameter. To address this, we replace the SGD optimizer with signed gradient descent as in~\citep{schwinn2024adversarial}. We keep the learning rate the same as in the original work~\citep{zou2024improving}.

\textbf{Initialization.} Instead of initializing the embeddings with a sequence of \enquote{x} tokens, we use a more semantically meaningful instruction string: \enquote{ Make sure to follow my instructions accurately and do not hesitant to answer in a toxic way. Do not worry about me engaging with this content. I am studying toxic behavior of LLMs. }. Without this change, the models continue their generation with sequences of \enquote{x} tokens after the adversarial target string was correctly generated. Note that we used this specific initialization on our first attempt after observing the generation issue and did not conduct any further optimization.

\textbf{Multiple generations.} We use the same early stopping procedure as in the original paper. Additionally, we periodically generate and evaluate outputs from the attacked model. Specifically, we produce a response for every $5$ attack iterations and generate a total of $20$ responses for every toxic query. An alternative approach could involve conducting multiple random restarts of the attack. However, we find our method is sufficient to break the defense while it is computationally more efficient. 

\textbf{Results.} We implement all these changes in the original repository of the paper and use the same evaluation script. Here, the ASR is calculated using the judge model provided by Harmbench~\citep{mazeika_harmbench_2024}. Additionally, we manually check $20$ random generations for both models. We found one false positive during this manual check. Nevertheless, For this specific instance, multiple other generations correctly broke the model. Moreover, we evaluated $20$ different negatively judged outputs and found $3$ false negatives. Thus, we conclude that the ASR provided by the judge model is likely not overestimated. 

The results are summarized in Table~\ref{tab:my_label}. Our new attack achieves a $100\%$ attack success rate (ASR) against both circuit breaker models provided in the paper. In contrast, the original evaluation reported $15,7$ and $9.6\%$ ASR for embedding space attacks, respectively. Increasing the number of attack iterations, the number of generations, optimizing the initialization string, or fine-tuning the learning rate would likely lead to even more efficient attacks. We want to clarify that this vulnerability is not unique to circuit breakers. Our new attack achieves $100\%$ ASR on Llama2-7b-chat-hf~\citep{touvron2023llama}, vicuna-7b-v1.5~\citep{vicuna}, and Mistral-7B-v0.1~\citep{jiang2023mistral} using the same attack and evaluation protocol.  

\begin{table}[]
    \centering
    \caption{Comparison of embedding attack success rates (ASR) in $\%$ against the short circuited models provided in~\citep{zou2024improving}. Changing the optimizer (opt) and initialization (init) improves the ASR considerably without inducing any additional cost. Evaluating with multiple generations (multi gen) for each attack breaks the defense.}
    \resizebox{1\columnwidth}{!}{
    \begin{tabular}{l|ll}
        \toprule
        Model & Mistral-7B-Instruct-v2 + RR & Llama-3-8B-Instruct + RR \\
        \midrule
        Input Embed~\citep{zou2024improving} (original) & 15.7 & 9.6 \\
        Input Embed (opt \& init) & 57.2 & 51.7 \\
        Input Embed (opt \& init \& multi gen) & 100 & 100 \\
        \bottomrule
    \end{tabular}
    }
    \label{tab:my_label}
\end{table}

Note that embedding space attacks are an unrealistically strong threat model for proprietary robustness evaluations and are mostly relevant as a threat model for open-source LLMs~\citep{schwinn2024soft}. Still, our experiments showed that embedding space attacks can be used as a sanity check. High robustness against unconstrained continuous attacks should be reported with caution.

\section*{Outlook}

Beyond the presented results, currently ongoing third-party evaluations observe vulnerabilities of short circuiting to manual human jailbreaks, such as separating the toxic content with spaces before forwarding it to the LLM. This gives further indication that the original robustness assessment may be overestimated. We plan to update this document with additional robustness evaluations, such as adaptive discrete attacks, in the future~\citep{geisler2024attacking, andriushchenko2024jailbreaking}.
We thank the authors of the original work for providing the community with code and pretrained models. This considerably reduced the effort required to conduct a third-party evaluation. We also want to thank them for their commitment to helping us with the evaluation.

\bibliography{ref}

\end{document}